\documentclass[twocolumn,preprintnumbers,amsmath,amssymb]{revtex4}
\usepackage{graphicx}
\usepackage{dcolumn}
\usepackage{bm}

\begin{document}
\title{Origin of $n$- and $p$-type conductivity in undoped $\alpha$-PbO: role of defects}
\author{J. Berashevich, J.A. Rowlands, A. Reznik}
\affiliation{Thunder Bay Regional Research Institute, 290 Munro St., Thunder Bay, ON, P7A 7T1, Canada}
\affiliation{Department of Physics, Lakehead University, 955 Oliver Road, Thunder Bay, ON, P7B 5E1, Canada}

\begin{abstract}
First principles calculations (GGA) have been applied to study
the crystallographic defects in $\alpha$-PbO in order to understand
an origin of $n$- and $p$-type conductivity in otherwise undoped $\alpha$-PbO.
It was found that deposition in oxygen-deficient environment to be defined in our simulations
by the Pb-rich/O-poor limit stimulates a formation of O vacancies
and Pb interstitials both characterized by quite low formation energies
$\sim$ 1.0 eV. The O vacancy, being occupied by two electrons, shifts the
a balance of electrons and holes between these two defects to an excess of electrons
(four electrons against two holes) that causes $n$-type doping.
For the Pb-poor/O-rich limit, an excess of oxygen
triggers the formation of the O interstitials characterized by such a low
formation energy that spontaneous appearance of this defect is predicted.
It is shown that the concentration of O interstitials is able to reach an extreme magnitude equal
to number of the possible defect sites ($\sim 10^{22}$cm$^{-3}$).
The localized state formed by the O interstitial is occupied by two holes and because
there are no other defects in reasonable concentration to balance the hole redundancy, $p$-type doping is induced.
\end{abstract}

\maketitle
Recent developments in radiation medical imaging technology impose new requirements 
on photoconductive materials: 
polycrystalline and amorphous materials are considered for 
application in flat panel detectors \cite{simon} with preference given to
high Z (atomic number) polycrystalline semiconductors. Among all polycrystalline materials, 
Lead Oxide ($\alpha$-PbO) is the most promising candidate due to its
long history of employment in optical imaging devices \cite{book}. 
However, despite such extensive application, properties of $\alpha$-PbO
are not well established. There are limited data on the transport properties: 
the low mobility-time product
indicates poor carrier transport \cite{rowland} and, therefore, 
it is believed that carrier transport should be controlled by trapping on defects.
For application of PbO in real time procedures such as fluoroscopy, 
transport characteristics are required to be optimized, 
for which an understanding of the nature of the defects is important.
The thermal evaporation technique used for deposition 
of PbO layers \cite{simon,book} presumes a 
deficiency of atomic oxygen ultimately leading to 
formation of oxygen vacancies. 
The conductivity enhancement of PbO samples observed after thermal annealing in an
atmosphere of oxygen supports this hypothesis \cite{chienes}. 
In our previous studies devoted to defect formation in $\alpha$-PbO single crystal,
we confirmed that this compound tends to
loose oxygen due to the low formation energy of the oxygen vacancy \cite{defects}.
Oxygen vacancies are found to be occupied by two electrons that leads to a prediction 
of excess of electrons. The 
experimentally observed $n$-type conductivity of the PbO layers 
grown under conditions of oxygen deficiency \cite{book,bigelow,chienes,scanlon} verifies
our theoretical findings \cite{defects}.

Samples characterized by excess oxygen show the exact opposite behavior: 
$p$-type conductivity \cite{bigelow}.
The Pb vacancies considered in our previous study \cite{defects}, although they are
occupied by two holes and the oxygen rich conditions promote their formation,
can not be responsible for $p$-type doping because of their rather high formation energy.
In order to develop the post-growth annealing procedures 
directed to suppression of the vacancy defects, 
an understanding of whether the origin of $p$-type conductivity is defect related 
is urgently required: the application of
annealing in oxygen atmosphere becomes questionable, provided it 
generates new defects, while healing the oxygen vacancies.
Therefore, to obtain a real picture a comprehensive study 
of all type of defects that can be formed under different growth conditions is required. 
Here we apply first-principles methods for such a study
(the Pb-rich/O-poor and Pb-poor/O-rich limits of the growth conditions are considered) with focus 
on an analysis of the defects being inherent to the $\alpha$-PbO crystal structure,
the appearance of the defect states inside the band gap, charged states and formation energies.

\section{Methods}
\label{sec:methods}
We study the formation of the defects in $\alpha$-PbO with help of the density functional theory (DFT)
provided by the WIEN2k package \cite{wien} which adopts 
the full-potential augmented plane-wave + local orbitals method. 
The generalized gradient approximation (GGA) \cite{GGA}
with the Perdew-Burke-Ernzerhof parametrization was used.
The $5p$, $5d$, $6s$ and $6p$ electrons of the Pb atom and 
$2s$ and $2p$ electrons of the O atom have been treated as valence electrons, 
while lower energy electrons were assigned to the core states (the energy cutoff was -8 Ry).
We applied a supercell approach with 108-atom size
(3$\times$3$\times$3 array of the primitive unit cells). 
The geometry optimization procedure applied to supercell containing a defect was performed based 
on minimization of the total energies and forces \cite{forces}.
The residual forces did not exceed 0.5 mRyd/Bohr, the energy convergence 
limit was set to 0.0001 Ry. The Brillouin zone was covered
by a 5$\times$5$\times$4 k-point mesh utilizing the Monkhorst-Pack scheme. 
The product of the atomic sphere radius 
and plane-wave cutoff in k-space was equal to 7.

The tetragonal lead oxide $\alpha$-PbO is characterized by a layered structure and
appears in polycrystalline form upon compound growth \cite{simon}.
The weak orbital overlap of the $6s^2$ lone pairs of Pb atoms projected out from each layer 
holds the PbO layers together \cite{new,new1}, while 
only $6p^2$ valence electrons participate in formation of the hetero-polar bonds with oxygen atoms \cite{new1}.
GGA tends to overestimate the interlayer separation in the layered structures while
underestimating the band gap size. 
For $\alpha$-PbO system in which the orbital overlap between layers controls 
the size of the band gap, it results in a compensation effect \cite{defects}. 
In particular, for the lattice parameters optimized 
with GGA ($a_0$=4.06 \AA\ and $c_0$=5.51 \AA\ \cite{defects}), the band gap is only slightly 
underestimated, as 1.8 eV to be compared with the experimental value of 1.9 eV \cite{exp} 
(for this compound, good agreement of the size of the band gap estimated theoretically with that found 
experimentally
has been pointed out in earlier work as well \cite{new}).
Application of the experimental value of the lattice parameters causes the band gap to shrink 
by 0.22 eV \cite{defects}. Because the size of the band gap is crucial in order to 
correctly define the appearance of the defect states inside the band gap, 
the band structure properties
have been evaluated for the optimized lattice parameters.
However, for calculations of the defect formation energy, an 
adjustment of the interlayer distance 
by using the experimental data of the lattice constant ($c_0=5.07$ \AA \cite{venk}) has been performed.
It was found that only the formation energy of the Pb interstitial  
is considerably affected by the mismatch in the lattice parameters, caused 
by the poor incorporation of Pb atom of large atomic radius between the layers
(the formation energies of the vacancy defects have been pointed out not to be 
affected by GGA limitation in our previous work \cite{defects}). 

\section{Crystallographic defects in $\alpha$-PbO}
Analysis of the literature on defect formation in semiconductors \cite{walle} and particularly
in simple oxides \cite{Kotomin}, has revealed that additionally to the vacancy defects investigated in our previous study \cite{defects},
interstitials have to be considered. 
For crystalline materials, interstitial defects
induce a large lattice distortion into their immediate neighbourhood that results
in their high formation energy \cite{walle}. In this regard, 
the crystal lattice of polycrystalline $\alpha$-PbO is unique,
because the distance between weakly interacting layers 
is large enough to accommodate a foreign atom ($c_0=5.07$ \AA \cite{venk}).
Therefore, along with the vacancy defects, formation of the interstitials is expected to be inherent to the
$\alpha$-PbO lattice.

\begin{figure}
\includegraphics[scale=0.19]{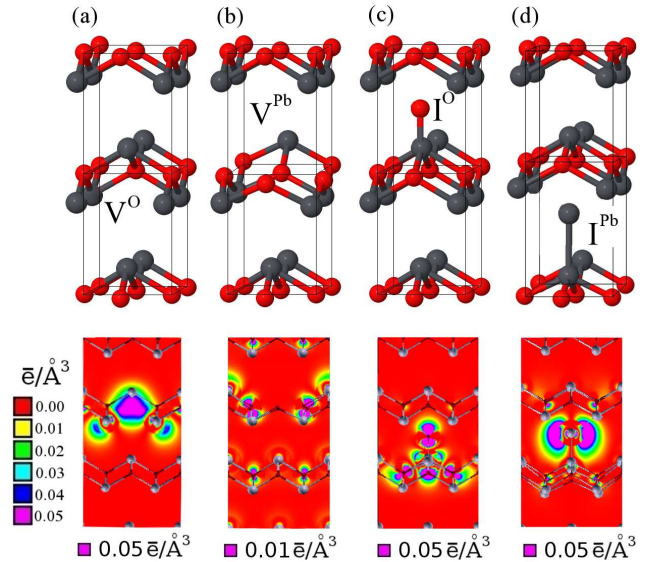}
\caption{\label{fig:fig1} (Colour on-line) Schematic representation of
crystallographic defects in the 
tetragonal $\alpha$-PbO (space group 129P4/nmm) and the 
electron density map plotted with 
XcrySDen \cite{xcrysden} for the energies $E_{D}\pm 0.05 eV$.
(a) the O vacancy, (b) the Pb vacancy, 
(c) the O interstitial and
(d) the Pb interstitial. The electron density maps (e/\AA$^3$)
are plotted with XcrySDen. 
The colour scale on the left side is for the 
$V^{\operatorname{O}}$, $I^{\operatorname{O}}$, $I^{\operatorname{Pb}}$, while 
for $V^{\operatorname{Pb}}$ it should be reduced by factor of 5).}
\end{figure}

\begin{figure}
\includegraphics[scale=0.38]{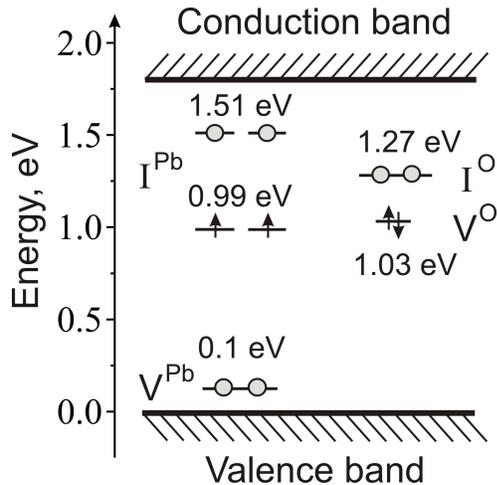}
\caption{\label{fig:fig2} Energetic location of the states induced 
by the various defects ($V^{\operatorname{O}}$, $V^{\operatorname{Pb}}$, $I^{\operatorname{O}}$, $I^{\operatorname{Pb}}$) 
within the band gap of $\alpha$-PbO ($E_G$=1.8 eV). Electrons occupying the vacancy site 
are indicated by the arrows while filled circles correspond to absence of 
electron.}
\end{figure}

We start from a short description of the vacancy defects already investigated in our previous work \cite{defects}. 
The O vacancy ($V^{\operatorname{O}}$) 
is occupied by two electrons localized on the 
four nearest Pb atoms with maximum of the electron density to manifest at the vacancy site 
(see the electron density distribution in Fig.~\ref{fig:fig1} (a)). 
The O vacancy can appear in three charged states (0, 1+, 2+). 
The localized state generated by the 
O vacancy is located deep inside the band gap
at $E_{D}$($V^{\operatorname{O}}$)-$E_{V}$=1.03 eV, where $E_{V}$ is the 
top of the valence band. 
In contrast, the Pb vacancy creates a localized state occupied by two holes and
the defect level appears close to the valence band top at $E_{D}$($V^{\operatorname{Pb}}$)-$E_{V}$=0.1 eV.
The electron density is spread between O and Pb atoms showing long defect tails 
(see Fig.~\ref{fig:fig1} (b)) 
which is not surprising taking into account the
defect level appearance close to the valence band top (see Fig.~\ref{fig:fig2}).
The Pb vacancy has also three charge states (0, 1$-$, 2$-$) 
but its charge states would be of negative sign,
i.e. ionization of the vacancy occurs through an electron gain.
The effect of mismatch of the lattice parameters on location of the
vacancy-induced states was examined.
It is found that a position of the defect states relatively to the valence band top
is not affected with application of the experimental lattice parameters:
since the vacancy states are localized entirely within single
layer, an alteration in the interlayer interactions has almost no impact on its property \cite{defects}. 

There are two interstitial defects to be considered 
in $\alpha$-PbO crystal lattice, the O and Pb interstitials. 
Binding of the
O interstitial ($I^{\operatorname{O}}$) to the host occurs 
with participation of the lone pair Pb:$6s^2$ electrons from the host
(the lone pair Pb:$6s^2$ in $\alpha$-PbO is chemically active \cite{new1}):
the bond is made between lone pair Pb:$6s^2$ electrons of the host 
and O:$p^0_{z}$ empty orbital of oxygen atom. The O:$p^0_{z}$ empty orbital
appears upon excitation of the ground state O:$2p^2_{x}p^1_{y}p^1_{z}$ to the
O$^*$:$2p^4_{x+y}p^0_{z}$ state. The lowest energy state is achieved when
the O interstitial is positioned
just above the Pb atom (see Fig.~\ref{fig:fig1}(c), the bond length is 2.03 \AA).  
Inside the band gap, the localized state is located
above the midgap at $E_{D}$($I^{\operatorname{O}}$)-$E_{V}$=1.27 eV (see Fig.~\ref{fig:fig2}).
This localized state can accommodate up to two extra electrons 
and, therefore, there are three charge states (0, 1$-$, 2$-$) are assigned to it.

The energetically favourable position of the Pb interstitial ($I^{\operatorname{Pb}}$)
in the lattice is similar to the O interstitial, i.e. on top of the Pb atom.
The Pb interstitial is exited from the ground state of configuration 
Pb:$6s^2$$6p^2$ to the Pb$^*$:$6s^1$$6p^3$ state
thus allowing $6s^1$ and $6p^1_{z}$ electrons to contribute in bond formation with the host.
The host Pb atom participates in bonding, again with its Pb:$6s^2$ electrons (the bond length is 2.9 \AA).
The originated defect state is occupied by two electrons which are
Pb:$6p^2_{x+y}$ electrons belonging to $I^{\operatorname{Pb}}$ (see Fig.~\ref{fig:fig1} (d)).
Since each electron stays on its own orbital, Hund's rules dictate 
the ferromagnetic ordering of these electrons \cite{my_magn}. 
The spin-polarization energy defined as
the energy difference between the spin-unpolarized and spin-polarized states is found to be
$E_{pol}$=0.235 eV that indicates the stable triplet ground state \cite{my_magn}.
The Pb interstitial induces a localized state inside the band gap and spin-polarized solution separates 
the occupied and unoccupied states as
$E_{D}$($I^{\operatorname{Pb}}$)-$E_{V}$=0.99 eV and $E_{D}$($I^{\operatorname{Pb}}$)-$E_{V}$=1.51 eV (see Fig.~\ref{fig:fig2}), respectively. 
Ionization of the defect state formed by the Pb interstitial can occur 
as through electron gaining ((1$-$) for one electron added and 
(2$-$) for two electrons), so, through an excitation of existing electrons.
Overall, there are five charged states (2+, 1+, 0, 1$-$, 2$-$)
can be considered for the Pb interstitial.

Since properties of the interstitial defects are expected to depend on the 
interlayer distance, an alteration in position of the defect states inside the band gap with implementation
of the experimental lattice parameters has been investigated. It was found that for the 
O interstitial the defect state is shifted only by 0.05 eV outwards $E_V$. 
However, the effect is more pronounced for the Pb interstitial because 
of larger atomic radius of Pb atom. Thus, the state occupied by electrons
appears at $E_{D}$($I^{\operatorname{Pb}}$)-$E_{V}$=0.97 eV while empty state at
$E_{D}$($I^{\operatorname{Pb}}$)-$E_{V}$=1.25 eV. 
Moreover, for the defect state like Pb interstitial, 
DFT is known to delocalize the defect wave function and 
underestimate the splitting between
the unoccupied and occupied states that occurs due to self-interaction
of the unpaired electrons \cite{book1}. Therefore, defect wave-function
localization has been additionally examined with Hartree-Fock (HF) approach
applied directly to the unpaired electrons \cite{HF, book1} (in this way the
accuracy provided by DFT is preserved). Indeed, it resulted 
in stronger localization and, therefore, enhanced splitting of the defect states 
such as 
$E_{D}$($I^{\operatorname{Pb}}$)-$E_{V}$=0.54 eV for the occupied state and
$E_{D}$($I^{\operatorname{Pb}}$)-$E_{V}$=1.66 eV for the empty state. 
However, since the experimental lattice parameters and HF correction
have opposite effect on the defect wave function localization, application of
both leads to compensation effect in level splitting such as appearance of the 
defect states in the band gap is 
$E_{D}$($I^{\operatorname{Pb}}$)-$E_{V}$=0.54 eV for the occupied state 
and $E_{D}$($I^{\operatorname{Pb}}$)-$E_{V}$=1.22 eV for the empty state.
Since the band gap size is reduced by 0.22 eV \cite{defects}, location of these
defect states in respect to bottom of the conduction band is very close to
that found with GGA (see Fig.~\ref{fig:fig2}). 

We have established that some defects can act as a
donor impurity ($V^{\operatorname{O}(0)}\rightarrow V^{\operatorname{O}(1+/2+)}$), others 
as acceptors ($V^{\operatorname{Pb}(0)}\rightarrow V^{\operatorname{Pb}(1-/2-)}$ and 
$I^{\operatorname{O}(0)}\rightarrow I^{\operatorname{O}(1-/2-)}$), 
while $I^{\operatorname{Pb}}$ can be both an electron donor 
($I^{\operatorname{Pb}(0)}\rightarrow I^{\operatorname{Pb}(1+/2+)}$) and an
acceptor ($I^{\operatorname{Pb}(0)}\rightarrow I^{\operatorname{Pb}(1-/2-)}$). 
An ability of defects to appear in several charge states encourages
the electron exchange between defects: the
electron donor compensates missing electrons on the acceptor.
The process of the charge exchange defines the energetically favorable charge state for defects
appearing upon compound growth.

\section{Formation energy of defects}
The main way to identify formation of defects inherent to the crystal structure
is to compare the formation energies of different defects, as particularly this parameter
is a measure of the defect concentration. 
The formation energy of a defect $D$ in charge state $q$ (+2/+1/0/-1/-2) can be defined as \cite{walle}
\begin{equation}
\Delta E^f(D)=E_{tot}(D^q)-E_{tot}(S)-\sum_{i}n_{i}\mu_{i}+q(E_F+E_V+\Delta V)
\label{eq:one}
\end{equation}
where $E_{tot}(D^q)$ and $E_{tot}(S)$ 
are the total energy of the system containing the single defect and 
defect-free system, respectively. 
$n_{i}$ indicates a number of $i$-atoms removed ($n_{i} > 0$) or added ($n_{i} < 0$), while
$\mu_{i}$ is the chemical potentials.
($E_F+E_V+\Delta V$) is the position of the Fermi level relative to
the valence band maximum ($E_V$) which has to be corrected by $\Delta V$ (for details see Ref. \cite{walle}). 

For calculation of the formation energy of the native point defects $\Delta E^f(D)$,
the supercell approach has been applied. In order to 
minimize an interaction between a defect and its periodic replicas, 
the sufficiently large supercell (3$\times$3$\times$3) is used for simulation of defects.
The geometry optimization procedure has been applied as for the neutral supercell 
(the total energy of the system corresponds to $E_{tot}(S)$) 
and so after defect has been charged: 
$E_{tot}(D^q)$ energy has been evaluated for the 
defects in the different charge states $q$ (+2/+1/0/-1/-2) (for details see \cite{walle}).
An alignment of the top of the valence band ($E_V$) for supercell with charged defect 
to that in the bulk is performed with help of the correction term $\Delta V$ \cite{walle}.
Since for accurate calculation of the formation energy of interstitial defects 
the mismatch in the interlayer distance ($c_0$=5.51 \AA) occurring
for the lattice parameters optimized with GGA in comparison 
to the experimental data ($c_0$=5.07 \AA \cite{venk}) is essential, 
the experimental value has been used for those calculations.
In this case, an application of the so-called band gap error $\Delta E_G$ \cite{togo}
is required to compensate for a reduction of the band gap size \cite{defects}.
For vacancies, because of their weak interactions with the opposite layer,
the formation energy would be rather insignificantly affected by
the interlayer distance mismatch, and the same argument works 
for the vacancies appearing on the surface of the single crystal (platelet).
However, for interstitial defects because of their
strong interaction with the opposite layer, the formation energy would be altered
upon appearance on the platelet's surface.
For defects on surface, slab’s thickness was four layers, 
its width was $\sim$10 \AA, while distance between slabs was $\sim$14 \AA.

To simulate the deficit and excess of oxygen, the 
Pb-rich/O-poor and Pb-poor/O-rich limits have been applied in evaluation of the 
chemical potentials. The chemical potentials have been found for the Pb-rich/O-poor limit as
$\mu_{\operatorname{(O)}}$=$E_{tot}$(O atom)+$\Delta_f H^0(\operatorname{PbO})$, where $\Delta_f H^0(\operatorname{PbO})$ 
is the standard enthalpy of formation of PbO compound, 
while $\mu_{\operatorname{(Pb)}}$ is determined 
from $\Delta_f H^0(\operatorname{PbO})$=$\mu_{\operatorname{(O)}}$+$\mu_{\operatorname{(Pb)}}$.
For the Pb-poor/O-rich limit the chemical potentials are
$\mu_{\operatorname{(Pb)}}$=$E_{tot}$(Pb atom)+$\Delta_f H^0(\operatorname{PbO})$ 
and, similarly, $\mu_{\operatorname{(O)}}$ is found from 
$\Delta_f H^0(\operatorname{PbO})$=$\mu_{\operatorname{(O)}}$+$\mu_{\operatorname{(Pb)}}$.
The standard enthalpy of formation $\Delta_f H^0(\operatorname{PbO})$=-2.92 eV per Pb-O pair has been 
calculated elsewhere \cite{defects} applying a formalism described in details earlier \cite{walle}.

\begin{figure}
\includegraphics[scale=0.50]{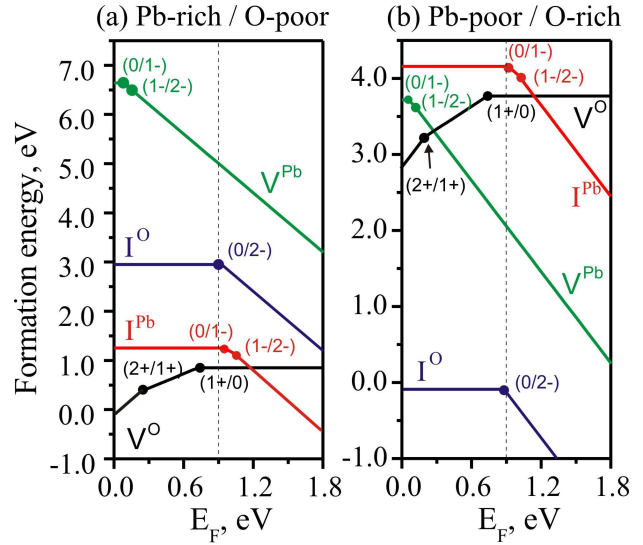}
\caption{\label{fig:fig3} Formation energy as a function of the Fermi level position 
($E_F$ varies from 0 to $E_G$=1.8 eV \cite{masses}). (a) Oxygen deficit (Pb-rich/O-poor limit). 
(b) Oxygen excess (Pb-poor/O-rich limit).}
\end{figure}

We have calculated the formation energies of the different defects in the relevant charge states 
and our results as a function of the Fermi level position for the 
Pb-rich/O-poor and Pb-poor/O-rich limits are shown in Fig.~\ref{fig:fig3}.
Since the Fermi level defines a probability for electron 
to occupy the defect state, the formation energy of the charged state 
is a function of Fermi level position inside the band gap.
As $\mu_i$ deviates in a range defined by $\Delta_f H^0(\operatorname{PbO})$ (in real growth conditions to be controlled 
by the partial pressure of reactants) it induces a shift of the formation energies.
The concentration of the defects can be estimated through their formation energies
as $N_{defects} \sim N_{sites}$ exp$(-\Delta E^f(D)/k_bT)$, 
where $k_b$ is the Boltzmann constant, $T=570$ K is the deposition temperature 
and $N_{sites}$ is the concentration of sites available for defects to be formed. 

For the Pb-rich/O-poor limit (see Fig.~\ref{fig:fig3} (a)), the formation energies 
of the O vacancy ($V^{\operatorname{O}}$) and 
the Pb interstitial ($I^{\operatorname{Pb}}$) appear
to be much lower than that for other defects and, 
therefore, these defects are expected to dominate.
The defect concentrations estimated for their zero charge states are
$10^{15}$cm$^{-3}$ for the O vacancy and $10^{11}$cm$^{-3}$ for the Pb interstitial.
For the Fermi level assigned to the midgap, 
the (0, 1+) and (0, 1-) states are energetically preferable for the 
O vacancy and the Pb interstitial, respectively.
Since the thermodynamic transition $V^{\operatorname{O}(0)} \rightarrow V^{\operatorname{O}(1+)}$
is found to occur almost at the midgap, it indicates a 
potential for both states to be formed during the compound growth.
However, to be ionized, $V^{\operatorname{O}(0)}$ has to dispose of its electron and therefore,
the appearance of the $V^{\operatorname{O}(1+)}$ charge state would naturally depend on the presence of the 
electron acceptor. 
For the Pb interstitial, we have calculated the formation energy for the 
five states (2+, 1+, 0, 1$-$, 2$-$). Since the formation energy of the (1-) state that occurs when an extra electron 
is added, is significantly lower than that when an electron is removed 
($\Delta E^f(I^{\operatorname{Pb}(1+)})$=3.98 eV for $E_F$ at midgap), 
the $(1+)$ and $(2+)$ states have been disregarded. 
The thermodynamic transition 
$I^{\operatorname{Pb}(0)} \rightarrow I^{\operatorname{Pb}(1-)}$
occurs close to the midgap and, therefore the charge states (0, 1$-$)
appear to be the preferable states. To become ionized to the (1$-$) state, 
$I^{\operatorname{Pb}(0)}$ has to gain an electron (electron acceptor).

Therefore, the charge exchange between $V^{\operatorname{O}(0)}$ and $I^{\operatorname{Pb}(0)}$ 
is possible as one of them can work as an electron donor while other as an acceptor.
The concentration of ionized states to manifest upon charge exchange would be limited 
by the defect of lower concentration, i.e. by the Pb interstitial ($10^{11}$cm$^{-3}$).
It is important to emphasize that the balance of electrons and holes occupying 
both defect sites in either combination of the charge states 
($V^{\operatorname{O}(0)}$/$I^{\operatorname{Pb}(0)}$ or 
$V^{\operatorname{O}(1+)}$/$I^{\operatorname{Pb}(1-)}$) is always shifted 
to a higher population of electrons (four electrons against two holes). 
Therefore, with respect to the conductivity type,
redundant electrons induced by the O vacancies 
would cause $n$-type doping of $\alpha$-PbO at the Pb-rich/O-poor limit to be in agreement 
with our previous studies \cite{defects}.
In the non-equilibrium electron gas (photogenerated or injected carriers), the ionized O vacancies 
$V^{\operatorname{O}(1+)}$ and the Pb interstitials in both charge states 
$I^{\operatorname{Pb}(0)}$, $I^{\operatorname{Pb}(1-)}$
would work as trapping centers for electrons in the conduction band. 

For the Pb-poor/O-rich limit, only the O interstitial, $I^{\operatorname{O}}$, 
possesses a low formation energy, while the formation energies of other
defects ($I^{\operatorname{Pb}}$, $V^{\operatorname{O}}$ 
and $V^{\operatorname{Pb}}$) are higher than 3.0 eV. 
In fact, the formation energy of $I^{\operatorname{O}}$ being so closed to zero 
implies the spontaneous appearance of this defect such as the defect concentration 
can become equal to the concentration of the possible defect sites ($N_{sites}\sim 10^{22}$cm$^{-3}$ \cite{Nsites}).
The O interstitial in its neutral charge state $I^{\operatorname{O}(0)}$
misses two electrons and, to be ionized to the charge states (1-/2-), has to acquire electrons. 
Since the thermodynamic transition 
$I^{\operatorname{O}(0)} \rightarrow I^{\operatorname{O}(2-)}$ occurs close to the midgap, 
$I^{\operatorname{O}}$ potentially can be formed in the charge states 
(1-,2-). However, because the concentration of centers of the donor type ($I^{\operatorname{Pb}}$, $V^{\operatorname{O}}$)
is negligible,
the O interstitial is expected to occur primary in its $I^{\operatorname{O}(0)}$ state.
The $I^{\operatorname{O}(0)}$ state would act as a trap for the non-equilibrium 
electrons in the conduction band. 
Therefore, under growth conditions corresponding to the oxygen excess, 
the high concentration of $I^{\operatorname{O}(0)}$ defects each to be occupied by two holes, 
would be able to cause $p$-type doping.
Moreover, the state formed by the O interstitial is characterized by a quite delocalized 
nature (see Fig.~\ref{fig:fig1} (c)): the long defect tails in combination with the high defect concentration 
would results in strong defect-defect interaction potentially leading 
to formation of the defect band.

It is also important to emphasize that all investigated defects, $V^{\operatorname{O}}$, $V^{\operatorname{Pb}}$,
$I^{\operatorname{O}}$ and $I^{\operatorname{Pb}}$ being considered in bulk $\alpha$-PbO
would be inherent to surface of the single crystal (platelet).
For some defects, the formation energy is found to be lowered 
when they are placed on the platelet surface.
The strongest effect is observed for 
the Pb interstitial for which the formation energy 
drops by $\sim$ 1.0 eV.
For the vacancies, the formation energy is only affected provided they appear at the edge of the platelet
(for example, it is reduced by 0.3 eV for oxygen vacancy).
The reduction of the formation energy at the platelet edge is 
a result of surface reconstruction leading to formation of the Pb-O double bonds instead of single bonds. 
Since the balance of electrons and holes occupying the defect sites 
is not affected by the formation of surface states, the considered 
phenomenon of the $n$- and $p$-doping remains valid.
There is one more interesting feature to be examined, which 
is defect pairing. For the Pb-rich/O-poor conditions, the pairing 
may occur between $V^{\operatorname{O}}$ and $I^{\operatorname{Pb}}$ defects, and is
promoted by a lowering of the formation energy by 0.84 eV. 
Such a pair $V^{\operatorname{O}}$-$I^{\operatorname{Pb}}$ still forms two defect levels at 
$E_{D}$($V^{\operatorname{Pb}}$)-$E_{V}$=1.22 eV and $E_{D}$($V^{\operatorname{Pb}}$)-$E_{V}$=0.66 eV 
and each level is occupied by two electrons.
For the Pb-poor/O-rich conditions, the interaction between $I^{\operatorname{O}}$ and $V^{\operatorname{Pb}}$
potentially can occur but because each localized state is occupied by two holes, such interaction 
is repulsive. Under equilibrium growth conditions,
the formation energy of $V^{\operatorname{Pb}}$ and $V^{\operatorname{O}}$ would become 
comparable, indicating a possibility of their pairing as well \cite{divacancy}. 
This combination of defects is unique as it induces vanishing of the dangling bonds
and, therefore, occurs without generation of a defect state inside the band gap.
Moreover, because the formation energy is lowered by 1.47 eV upon defect pairing,
it indicates the pairing to occur not only during the compound growth but also 
as a post-growth migration of the O vacancy towards the Pb vacancy.

\section{Conclusion}
Our study on the formation of defects in single crystal $\alpha$-PbO 
has revealed that some defects under certain growth conditions 
would appear in significant concentration causing $n$- or $p$-type doping of compound.
In particular, for the Pb-rich/O-poor limit
a formation of the $V^{\operatorname{O}}$ and $I^{\operatorname{Pb}}$ defects are found to dominate
with concentrations of
$10^{15}$cm$^{-3}$ and $10^{11}$cm$^{-3}$, respectively. 
The obtained defect concentrations are in good agreement with concentration of 
the localized states estimated experimentally ($10^{14}$-$10^{15}$cm$^{-3}$ \cite{book}).
It was established that accumulation of electrons on sites of the $V^{\operatorname{O}}$ and $I^{\operatorname{Pb}}$ defects
results in $n$-type doping of $\alpha$-PbO. 
For the oxygen excess created during the compound growth (Pb-poor/O-rich limit), another defect is found to manifest 
which is the O interstitial. The concentration of $I^{\operatorname{O}}$
is predicted to approach the highest limit because its formation energy is around zero, i.e.
the defect concentration may become equal to the number of the possible defect sites.
Since each $I^{\operatorname{O}(0)}$ is occupied by two holes, its high concentration 
would result in $p$-type doping of $\alpha$-PbO.
Moreover, such low formation energy of the O interstitial 
might be a problem for the post-growth procedures such as annealing.
We predict that annealing in an atmosphere of oxygen addressed to heal
O vacancies would generate new 
defects such as O interstitials.
With respect to the transport properties, 
the negative impact of the dominant defects, $I^{\operatorname{O}}$, 
$V^{\operatorname{O}}$ and $I^{\operatorname{Pb}}$ would become apparent
because these defects induce the states close to the midgap: severe trapping of
non-equilibrium electrons from the conduction band is expected.

\section{Acknowledgement}
Financial support of Ontario Ministry of Research and Innovation 
through a Research Excellence Program “Ontario network for advanced 
medical imaging detectors” is highly acknowledged.


\begin{thebibliography}{99}
\bibitem{simon}
Simon M, Ford R A, Franklin A R, Grabowski S P, Menser B,
Much G, Nascetti A, Overdick M, Powell M J and
Wiechert D U 2005 {\it IEEE Transactions on Nuclear Science} {\bf 52}, 2035.
\bibitem{book}
Photoelectronic imagining devices, V. 2 (1971), Eds. L.M. Biberman, S. Nudelman.
(Plenum Press, New York-London); Broek J 1967 {\it Philips. Res. Rep.} {\bf 22}, 367.
\bibitem{rowland}
Rau A W, Bakueva L and Rowlands J A 2005 {\it Med. Phys.} {\bf 32} 3160.
\bibitem{chienes}
Hwang O, Kim S, Suh J, Cho S and Kim K 2011 
{\it Nuclear Instruments and Methods in Physics Research A} {\bf 633} S69.
\bibitem{defects}
Berashevich J, Semeniuk O, Rubel O, Rowlands J A, Reznik A 2013 {\it J. Phys.:
Condens. Matter.} {\bf 25}, 075803.
\bibitem{bigelow}
Bigelow J E and Haq K E 1962 {\it J. Appl. Phys.} {\bf 33} 2980.
\bibitem{scanlon}
Scanlon D O, Kehoe A B, Watson G W, Jones M O, David W I F, 
Payne D J, Egdell R G, Edwards P P and Walsh A 2011 {\it Phys. Rev. Lett.} {\bf 107} 246402.
\bibitem{wien}
Blaha P, Schwarz K, Madsen G K H, Kvasnicka D and Luitz J {\it Wien2k: An Augmented 
Plane Wave + Local Orbitals Program for Calculating Crystal Properties: 
Karlheinz Schwarz}, (Techn. Universit\"at Wien, Austria, 2001)
\bibitem{GGA}
Perdew J P, Burke K and Ernzernof M 1996 {\it Phys. Rev. Lett.} {\bf 77} 3865.
\bibitem{forces}
Yu R, Singh D and Krakauer H 1991 {\it Phys. Rev. B} {\bf 43} 6411.
\bibitem{new}
Payne D J, Egdell R G, Law D S L, Glans P-A, Learmonth T, Smith K E, Guo J,  
Walsh A and Watson G W 2007 {\it J. Mater. Chem.} {\bf 17} 267.
\bibitem{new1}
Walsh A, Payne D J, Egdell R G and Watson G W 2011 {\it Chem Soc Rev} {\bf 40} 4455.
\bibitem{exp}
Thangaraju B and Kaliannann P 2000 {\it Semicond. Sci. Technol.} {\bf 15} 542.
\bibitem{venk}
Venkataraj S, Geurts J, Weis H, Kappertz O, Njoroge W K,
Jayavel R and Wuttig M 2001 {\it J. Vac. Sci. Technol. A} {\bf 19} 2870.
\bibitem{walle}
Van der Walle C G and Neugebauer J 2004 {\it J. Appl. Phys.} {\bf 95} 3851.
\bibitem{Kotomin}
Kotomin E A, Popov A I 1998 {\it Nucl. Instr. and Meth. in Phys. Res. B} {\bf 141}, 1.
\bibitem{xcrysden}
Kokalj A 2003 {\it Comp. Mater. Sci.} {\bf 28} 155.
\bibitem{my_magn}
Berashevich J and Reznik A, arxiv:1304.2945.
\bibitem{book1}
Advanced Calculations for Defects in Materials: Electronic Structure Methods (2011), Eds. 
Alkauskas A, Dek P, Neugebauer Jrg, Pasquarello A, Van de Walle, Chris G. 
(Weinheim, Germany: Wiley-VCH Verlag \& Co), Ch. 11.
\bibitem{HF}
d'Avezac M, Calandra M, Mauri F 2005 {\it Phys. Rev. B} {\bf 71}, 205210.
\bibitem{togo}
Togo A, Oba F, Tanaka I and Tatsumi K 2006 {\it Phys. Rev. B.} {\bf 74} 195128.
\bibitem{masses}
Berashevich J, Semeniuk O, Rowlands J A and Reznik A 2012 {\it EPL} {\bf 99} 47005.
\bibitem{divacancy}
Berashevich J, Rowlands J A and Reznik A 2013 {\it EPL} {\bf 102} 47002.
\bibitem{Nsites}
The standard procedure has been applied
to determine $N_{sites}$: 
$N_{sites}=N_{A}\rho/M$, where $N_{A}$ is the Avogadro's number,
$M$ is the molar mass and $\rho$
is the density found for $\alpha$-PbO at http://en.wikipedia.org/wiki/Lead(II)oxide.
\end{thebibliography}
\end{document}